# Fundamental mechanisms of hBN growth by MOVPE

**Krzysztof Pakuła, Aleksandra Dąbrowska, Mateusz Tokarczyk, Rafał Bożek, Johannes Binder, Grzegorz Kowalski, Andrzej Wysmołek, and Roman Stępniewski**

Faculty of Physics, University of Warsaw, Warsaw, Poland

krzysztof.pakula@fuw.edu.pl

## Abstract

Hexagonal boron nitride is a promising material for many applications ranging from deep UV emission to an ideal substrate for other two dimensional crystals. Although efforts towards the growth of wafer-scale, high quality material strongly increased in recent years, the understanding of the actual growth mechanism still remains fragmentary and premature. Here, we unveil fundamental growth mechanisms by investigating the growth of hBN by metalorganic vapor phase epitaxy (MOVPE) in a wide range of growth conditions. The obtained results contradict the widespread opinion about the importance of parasitic gas-phase reactions decreasing the growth efficiency. Two different growth mechanisms that depend on ammonia flow and reactor pressure can be distinguished. Both mechanisms are effective in the case of polycrystalline growth, but the growth of highly ordered, flat layers, is strongly hindered. The problem is caused by a low efficiency of boron chemisorption on N-terminated edges of sp²-BN sheets forming the atomic steps on the surface of the layer. Two-dimensional growth can be activated and sustained by the flow modulation epitaxy (FME) method, an alternate switching of ammonia and TEB flows. The success of the FME method is explained in terms of periodic changes between N- and B-terminated reconstructions at the edges of sp²-BN sheets, which restore boron chemisorption. The presented results identify the fundamental growth mechanisms, which is the prerequisite for any further deterministic development of efficient, high-quality, large-scale hBN growth with MOVPE.

## 1. Introduction

Hexagonal boron nitride (hBN) attracts increasing interest in many fields of physics and for many potential applications. hBN can be used in deep ultraviolet (DUV) optoelectronics [1], in flexible electronics [2], in thermionic energy conversion [3], as a host material for single photon emitters [4], insulating and barrier material for van der Waals heterostructures [5,6], the best suitable substrate for graphene and other two-dimensional crystals [7], a hyperbolic metamaterial [8], a material for neutron detectors [9], proton conductor [10], and pollutant remover [11]. Depending on the given application, different requirements are of importance, but most applications require uniformly grown layers on as large as possible substrates. This can only be satisfied by epitaxial technologies.

Hexagonal boron nitride consists of single atomic layers with sp² covalent bonds between the boron and nitrogen atoms giving rise to a typical honeycomb structure. The monolayers can interact with each other, via van der Waals forces, building a three-dimensional structure referred to as a van der Waals crystal [12]. In the case of a hexagonal boron nitride crystal, all boron atoms of one sp²-sheet are situated exactly underneath the nitrogen atoms of the next sp²-sheet and vice versa. An almost perfect hBN structure is observed in the case of small, bulk crystals obtained with high-pressure technologies [13]. In epitaxial films, sp²-sheets may be stacked differently, nominally destroying the hexagonal symmetry and forming rhombohedral (rBN) or turbostratic (tBN) systems [14]. In the presented work, the term "hBN" will be used in wide meaning, to describe a stack of sp²-BN monolayers.

Deposition of hBN layers is, most commonly, performed using chemical vapor deposition (CVD) techniques, with different precursors, on metallic substrates [15,16] or by using MOVPE on sapphire. One of the fundamental problems, reported for both techniques and all precursors and substrates, is a low efficiency of the growth. In many cases this results in an uncontrolled growth termination after formation of one, to a few, mono-atomic layers of hBN [17]. This, so called "self-terminated growth", is explained by differences between the physisorption energy on the substrate and hBN layer or, more generally, by a catalytic influence of the substrate that diminishes after the deposition of the first monolayer. Such an explanation is convincing in the case of metallic substrates and for lower temperatures [18], but cannot refer to all substrates and moreover not for thicker layers. Yet, a similar problem is reported in MOVPE, which means at high temperatures, on sapphire substrates, and even for hBN layers that are few monolayers thick [19,20].

MOVPE growth of hBN is typically performed with the use of triethylboron (TEB) and ammonia ($NH_3$) as boron and nitrogen precursors. Relatively thick, applicable layers, including p-type hBN:Mg [21], can be grown on a large scale [22]. Initial investigations of MOVPE growth, on a sapphire substrate, revealed a very low efficiency of BN synthesis. The problem was explained as the result of parasitic gas-phase reactions [23], in analogy with reactions between trimethylaluminium (TMAl) and ammonia, observed in aluminium nitride (AlN) epitaxy [24]. In the case of AlN the problem can be minimised by reactor design, lower ammonia flow and low reactor pressure (<200 mbar) [25]. In the case of hBN however, the quality of the layers deteriorates fast with decreasing ammonia flow [23]. The problem was successfully circumvented thanks to flow-rate modulation epitaxy (FME) [26]. An enhancement of the surface migration of ad-atoms was postulated to result from this method.

FME is based on an alternating switching of TEB and ammonia flows, which enables separation in time of the gaseous reagents. Previously it was implemented in aluminium nitride growth [27] as an expansion of atomic layer epitaxy. Recently, some basic aspects of the FME method, alternatively referred to as "pulsed growth" were presented including an empirical illustration of the basic dependencies on growth parameters [28,29]. Again, avoiding of parasitic gas-phase reactions and an improved diffusion on the growth surface were brought forward as explanation. Additionally, a problem concerning the uncontrolled formation of some kind of particles or islands on the layer surface was notified. The nature of the particles was again explained in terms of parasitic gas-phase reactions and insufficient diffusion.

All the above specified problems – the low growth efficiency, self-terminated growth, parasitic reactions, low diffusion of ad-atoms, mechanisms of FME growth, formation of disordered particles on the surface and, in addition, mechanism of generation of point-defects, responsible for optical properties of hBN, need to be addressed together in a wide, integral investigation. The explanations proposed to this date, regarding both, continuous flow growth (CFG) and flow modulation epitaxy (FME), remain essentially experimentally unexplored. The available information on growth conditions, are either inadequate or cover only a narrow range of growth conditions. The applicability of knowledge originating from the growth of "classical" covalent materials (GaN, AlN, GaAs) to the case of hBN growth was not verified, for example by specially designed experiments or studies covering a sufficiently wide range of conditions. Moreover, no detailed description of the MOVPE growth mechanisms for hBN, especially with regard to its van der Waals specificity, was proposed.

In this report we address the fundamental growth processes by performing a comprehensive study including a wide range of growth conditions. Most of the presented experiments were performed using CFG mode, which simplifies the interpretation of the results. The conclusions, drawn from CFG processes, were then used to develop and interpret also the results obtained using FME growth. In this report we will explain an obtained experimental trend in a stepwise manner which gradually merges into an overall picture describing the properties of the mechanisms governing the MOVPE growth of hBN.

## 2. Experiment

All growth processes were performed using an Aixtron CCS 3x2" MOVPE system designed for the growth of nitride compounds. Before the boron nitride growth processes, the system efficiency was calibrated using trimethylgallium (TMGa) and ammonia for gallium nitride (GaN) growth. The use of 100 μmol/min of TMGa resulted in a growth rate up to 35 nm/min. This value was used as a reference in BN growth efficiency investigations.

Boron nitride growth was performed on two-inch sapphire c-plane wafers, with 0,2º misorientation. Hydrogen, ammonia and TEB were used as the carrier gas and the precursors of nitrogen and boron, respectively. The reactor heating was controlled by a thermocouple, but the multipoint pyrometer was used to visualize the temperature distribution on the susceptor and the substrates. The growth temperature values given in this report are the substrate temperature obtained from the readout of the pyrometer system. The maximum temperature of the substrates, limited by the system design, was about 1300ºC.

The growth was monitored with a reflectometer based on a laser with 635 nm wavelength. In-situ reflectometry enables growth rate estimation for several different growth parameters values during one process run. It is a fast way to narrow down the parameter space and to identify interesting growth regimes. The laser beam is directed perpendicularly to the substrate surface and is reflected from the layer surface and from the interface between the layer and the substrate. Both reflected light beams interfere, leading to oscillations of the detected signal, which is a function of the increasing layer thickness. Each period of oscillations corresponds to the increment of the layer thickness for $D = \lambda/2n$, where $\lambda$ is the laser wavelength and $n$ is the refractive index of the layer material. In the case of hBN, the refractive index depends on the material orientation. For light with a polarization perpendicular to the c axis of the crystal, the refractive index reaches 2.27 [30,31]. For polarization parallel to c axis the refractive index decreases to 1.65 [30,31]. In the case of a polycrystalline, and porous material, the refractive index can be essentially lowered depending on porosity ratio. Additionally, during the growth, the layer structure can evolve, changing the effective optical properties. Due to these reasons the refractive index of polycrystalline BN layer can only be roughly estimated. The growth rate value calculated using such a parameter and its associated uncertainty is introducing a systematic error. However such an error still allows for a meaningful, though not quantitative, comparison of the main trends in growth rate changes as a function of basic process parameters.

The morphology and thickness of the layers was investigated ex-situ, using a Digital Instruments MMAFM-2 atomic force microscopy (AFM) and FEI Helios NanoLab 600 Dual Beam system (SEM). A Renishaw inVia system with a 532 nm continuous wave Nd-Yag laser excitation source was used for micro-Raman spectroscopy, to verify sp²-BN phase presence and layer uniformity. Far infrared reflectance was measured using a FT-IR microscope (Thermo

Fischer Scientific iS50, Nicolet Continuum) equipped with a 32x infinity corrected reflective objective (NA 0.65). Optical absorption measurements (Varian Cary 5000 UV-Vis-NIR Spectrometer) were used for the thickness estimation of the thinnest layers, below 10 nm. Photoluminescence (PL) was investigated with a Horiba T64000 spectrometer and a 532 nm laser source. The structural properties and thicknesses of the layers were also verified using X-ray diffraction (XRD) and X-ray reflectometry (XRR) with X'pert Phillips Diffractometer equipped with standard laboratory X-ray source (Cu K radiation) and a parallel beam Bragg reflection mirror.

## 3. Results and discussion

### 3.1 Growth rate determination methodology

A typical reflectivity signal collected during the continuous flow growth process with variable temperature is presented in Fig.1. The reactor pressure and ammonia flow were kept at P = 100 mbar and $NH_3$ = 100 ccm (~4500 μmol/min), which are a relatively low values compared to conventional nitride epitaxy. The gain of the initial reflectivity signal, coming from the substrate surface, was manually adjusted to R = 7.5%, to fit the sapphire substrate (n = 1.77). The process was started at T = 625°C with a flow of TEB = 40 ccm (~100 μmol/min). The Reflectivity was stable for 15 minutes, indicating growth failure. After increasing of the TEB flow to 80 ccm, the reflectivity signal started to increase and oscillate. At each oscillation maximum the growth was stopped, the temperature changed and the growth restarted. The increasing growth temperature led to a decrease of the oscillation period and a simultaneous increase of the amplitude. This relationship was confirmed with the last two process steps, involving large, contrary changes of the reactor temperature to 540°C and 990°C.

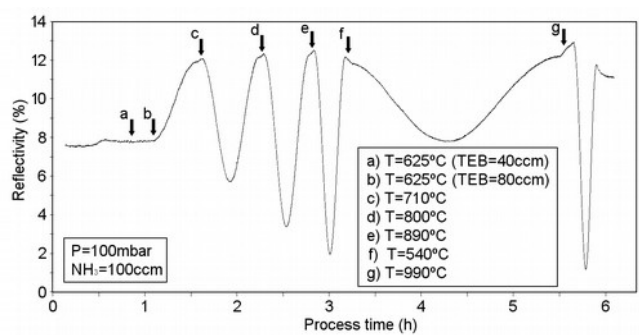

**Figure 1.** In-situ reflectivity of a BN layer, grown using the CFG mode, for different temperatures. The arrows indicate the moments, when the growth was interrupted and the parameters were changed.

During the growth, the maxima of the oscillating signal remained almost constant at a level of about 12%. This suggested the growth of optically flat, but polycrystalline, slightly porous, material with effective refractive index higher than for sapphire, close to n = 1.9. Assuming this value to be constant, the layer increment, corresponding to one oscillation period, was estimated to about D = 165 nm. The calculated growth rate value changed at the range of 1.2 to 11 nm/min, depending on temperature. These changes are well above any expected uncertainty coming from possible differences in the n value between the stages of the process.

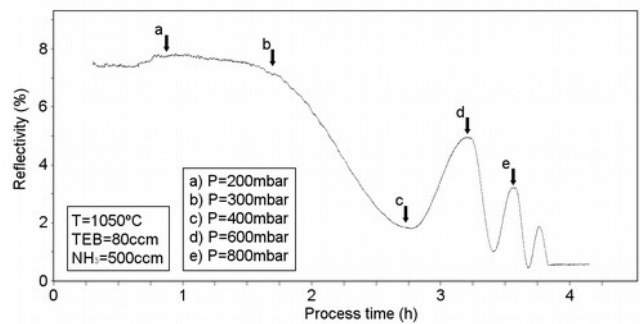

**Figure 2.** In-situ reflectivity of a BN layer, grown using the CFG mode, for different pressures. The arrows indicate the moments, when the parameters were changed.

Fig.2 shows the reflectivity signal recorded during the CFG process conducted at the temperature fixed at T = 1050°C, and ammonia flow $NH_3$ = 500 ccm, which is significantly more than in the previous case. The growth was started at the reactor pressure P = 200 mbar. The reflectivity decreased slowly from the beginning of the process, even when the amount of material was too low, to scatter the light. This indicates that the value of the effective refractive index was lower than for the sapphire substrate. This was the result of a highly porous layer structure. A successive increase of the reactor pressure led to an oscillation period shortening. The subsequent weakening of the optical signal can be explained by an increasing surface roughness and light scattering. The effective refractive index of the layer was assumed to be n = 1.6. Consequently, the layer thickness increment corresponding to oscillation period was estimated to D = 200 nm. With this assumption, the growth rate changes from 0.3 nm/min at P = 200 mbar to 16 nm/min at P = 800 mbar.

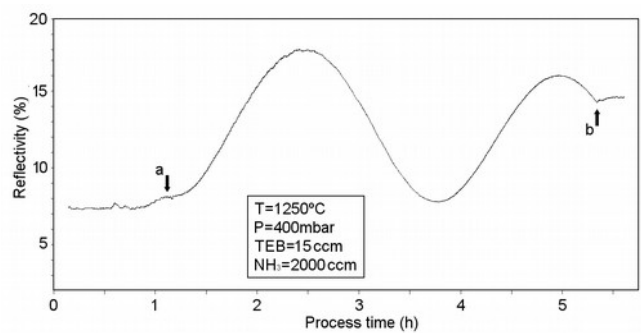

**Figure 3.** In-situ reflectivity of a BN layer grown using the FME mode. The arrows indicate the start and end of the growth.

The readout of the reflectivity value from the FME process is presented in Fig.3. Reflectometry from the BN layer shows an average value above 12%, matching a refractive index of n = 2.1, which indicates the presence of a solid material. The layer thickness increment corresponding to one oscillation period is D = 151 nm, resulting in a total thickness about 250 nm and a growth rate ~1 nm/min.

Measurements of the thickness, performed for several layers with SEM, AFM and XRR verified the estimations from reflectometry with a 20-30% accuracy. Discrepancies were largest for highly porous materials, where reflectometry provided underestimated values, corresponding rather to the mass than the volume of the porous layer. Despite of this inaccuracy, the reflectometry method proved reliable enough to compare layers grown under similar conditions. This fast method of the growth rate estimation was used for several processes, where the most important growth parameters were varied. The results, presented in the form of graphs, yield clear information about the kinetics of the growth process.

### 3.2 CFG processes with TEB supersaturation (3D growth)

The CFG process presented in Fig 1. indicated the necessity to use a large TEB flow to initiate growth. Most processes described in this section were performed with a TEB flow fixed at 80 ccm, ensuring a large growth rate up to 70 nm/min. The excessive amount of TEB, referred to as supersaturation, can lead to forced, disordered or non-stoichiometric growth of BN. On the other hand it can be necessary to counterbalance desorption from the growing surface or its decomposition.

The pyrolysis of TEB, like for many other metalorganic sources used in MOVPE, starts below 500ºC. In the range of temperatures used for the presented processes, we can assume fast and complete decomposition of TEB. Consequently we will in the following only discuss boron atoms, not TEB.

The dependence of the BN growth rate on the reactor pressure and the ammonia flow, using CFG mode, is presented in Fig.4 and Fig.5. The results were estimated from the reflectometry measurements. The processes were performed with fixed total gas flow. The ammonia flow parameter can be simply identified with ammonia partial pressure, but the total reactor pressure is a more complex parameter. Formally it is the sum of the partial pressures of all used gases, but a change of the total pressure, for a fixed total gas flow, leads also to the change of gas velocity, the time that reactants stay inside the reactor, the diffusion length in the gas phase and the desorption from the substrate surface.

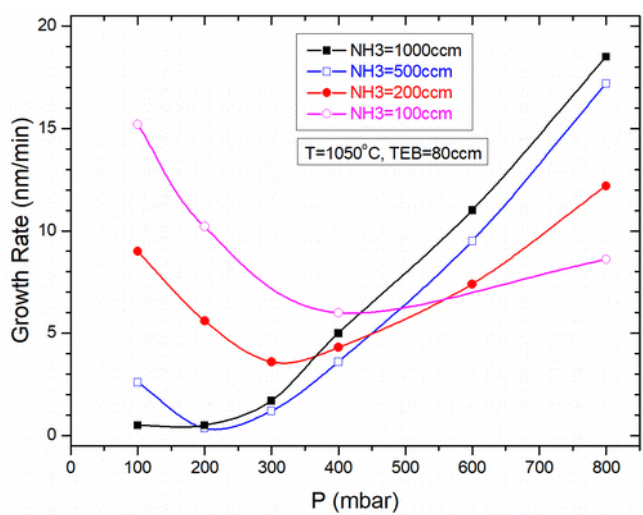

**Figure 4.** Growth rate dependence on the reactor pressure for different ammonia flows using the CFG mode. Lines are guides to the eye only.

The growth rate dependence on the total pressure (Fig.4) shows different trends for different ammonia flows. The simplest trend is observed for the largest ammonia flow ($NH_3$ = 1000 ccm). The synthesis efficiency is very low at the lowest pressure (P < 200 mbar), but increases fast with an increase in reactor pressure. This finding strongly contradicts suggestions, that BN synthesis can be limited by parasitic reactions in the gas phase, similar to that observed for AlN. In such a case, a higher reactor pressure should increase the probability of gas-phase reactions as well as the creation of volatile boron compounds and should lead to a diminishing synthesis efficiency [25]. However, Fig. 4 shows the exact opposite trend. For the same reasons, the gas-phase reaction probability should increase with increasing ammonia amount and process temperature, which is not observed (Fig.5 and Fig.6). It can hence be concluded, that parasitic gas-phase reactions are not of major importance for BN synthesis with ammonia and TEB precursors.

The hypothesis about the presence of such reactions was postulated on the basis of dependencies observed only at low (< 200 mbar) total reactor pressure, not taking into account the strong increase observed for larger values. Indeed, in the case of a low ammonia flow ($NH_3$ < 500 ccm), the growth rate increases for lower pressures. This dualism is even better visible in Fig.5, where the dependencies on ammonia flow for lowest and highest pressures show opposite trends.

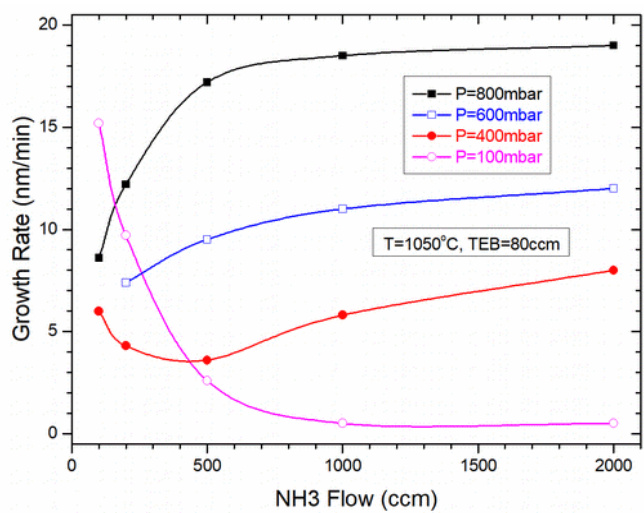

**Figure 5.** Growth rate dependence on the ammonia flow for different reactor pressures using the CFG mode. Lines are guides to the eye only. The corresponding V/III ratio value was changed in the range of 24 to 480.

We can hence identify two growth regimes, for which the fastest growth is observed: (i) with high ammonia flow at high reactor pressure (HAHP) or (ii) with low ammonia flow at low reactor pressure (LALP). These two types of conditions can be roughly identified as nitrogen-rich and nitrogen-poor, respectively. Such a successful growth under largely different conditions indicate the existence of two different mechanisms governing the BN synthesis. In order to better understand the growth mechanisms, the dependence on the growth temperature was extended to temperatures up to T=1300ºC. Results for HAHP and LALP conditions, presented in Fig.6, follow two independent trends, which constitutes further strong evidence for a full separation of the two different synthesis mechanisms.

The growth under LALP conditions can be observed starting from T = 550ºC, which is comparable with other

MOVPE reactions and results from a low thermal stability of metalorganic compounds. The activation of ammonia, as nitrogen source, is the result of a heterogeneous pyrolysis on the surface [18] under conditions of relative excess of boron. An important observation is that the synthesis efficiency increases slowly and almost linearly with the reactor temperature and does not saturate within the investigated range. The highest observed growth rate, about 25 nm/min, stays clearly below the maximal value expected for this particular TEB flow. By linear extrapolation, the saturation could be expected at much higher temperatures, even above 1500ºC.

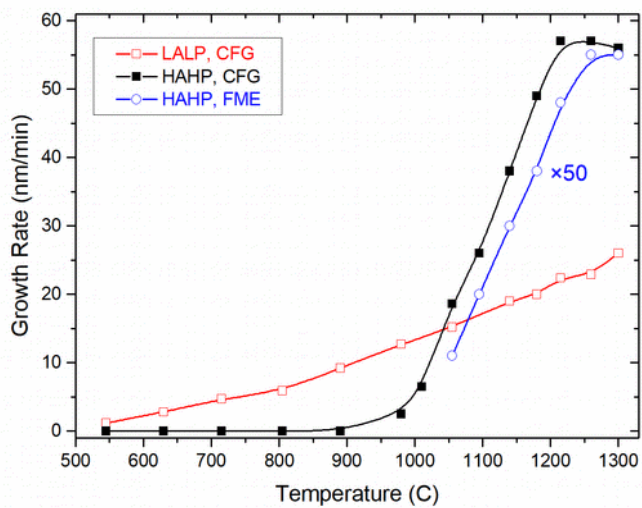

**Figure 6.** Growth rate dependence on temperature for different growth modes and conditions. Continuous flow growth (CFG) was investigated at LALP and HAHP conditions with TEB = 80 ccm. FME processes were performed at HAHP conditions with TEB = 15 ccm. Lines are guides to the eye only.

The linear dependency on the temperature implies that no activation energy is limiting this growth mechanism under LALP conditions. The observed dependency has to be the joint result of some physical processes, like physisorption, desorption and diffusion of boron and ammonia on the growing surface. Taking into account the strong dependence of the growth efficiency on the amount of ammonia at the lowest pressure (Fig.5) one may speculate that a higher temperature (Fig.6) increases the ammonia desorption, additionally diminishing the ammonia amount, which increases the heterogeneous catalysis efficiency. The fact that hBN can be successfully grown at temperatures as low as 550ºC is very important for future applications, as it enables direct hBN growth on substrates that cannot sustain high temperatures. This fact can prove important for possible future complex III-V epitaxy including hBN.

Layers grown at LALP conditions are mirror-like, transparent, slightly yellowish for the naked eye and consist of dense grains, below 100 nm in diameter, as shown in Fig.7. The grains do not reveal any symmetry, but the shapes are rather angular with sharp edges. The small grain sizes suggests a low diffusion length on the surface and a dense, repeated, disordered nucleation of new grains. But it can be interpreted, more generally, as growth in variable directions.

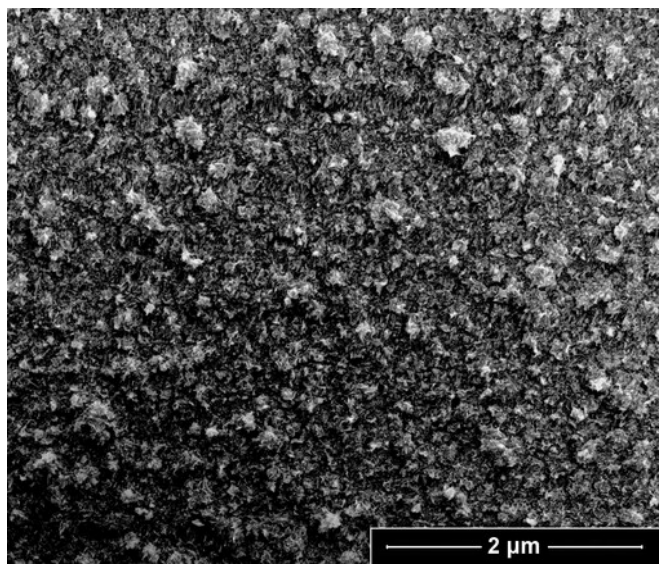

**Figure 7.** SEM image of a BN layer (sample A) grown under LALP conditions at T = 1050ºC. The average thickness was estimated from reflectometry to 380 nm.

Growth under HAHP conditions can be observed only above T=950ºC (Fig.6). The efficiency increases relatively fast with increasing temperature and saturates above 1200ºC. The highest observed value, about 60 nm/min, is close to the expected maximal value estimated by comparison with GaN growth. This behaviour indicates an existence of a well-defined energetic barrier (activation energy) characteristic for chemical reactions. A possible candidate could be the ammonia decomposition similar to that observed above 900ºC in the presence of Ga and at GaN surface [32]. According to terminology used by G.B. Strinfellow, the growth of BN below 1200ºC, at HAHP conditions, should be classified as kinetically limited growth [33]. Above 1200ºC the ammonia is decomposed fast and completely, which should enable control of the growth by mass transport only.

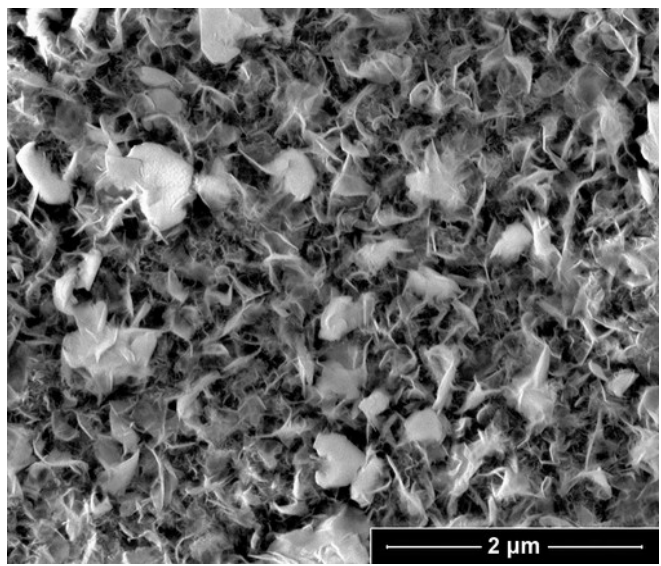

**Figure 8.** SEM image of a BN layer (sample B) grown under HAHP conditions at T=1050ºC. The average thickness was estimated from reflectometry to 380 nm.

Layers grown under HAHP conditions are matt and colourless for the naked eye. The SEM reveals (Fig.8) their structure to be polycrystalline, consisting of grains in the form of thin flakes with diameters comparable to the average thickness of the layer. Most of the visible flakes are arranged more or less perpendicularly to the substrate surface, which renders the layer porous. Such a shape of the grains indicates a highly anisotropic growth, with very effective diffusion on the $sp^2$-BN sheet surface and chemisorption acting mostly at the edges of the sheets, which is the only area with unsaturated covalent bonds. Most of the flakes are bent or even convoluted indicating high flexibility.

Occasionally, for layers grown at HAHP conditions, grains oriented more parallel to the substrate can be observed (Fig.9). They can reach diameters above 1 μm, larger than average thickness of the layer. Such pyramid-like grains show angular shapes, with flat walls. A closer look reveals the tendency to creation of pentagonal symmetry of the pyramids. Similar five-fold symmetry was described for hBN nanoparticles as a stack of monolayers folded around central nitrogen vacancies [34]. Each monolayer consist of five $sp^2$-BN sheets forming pyramid walls. Linear boundaries between misoriented $sp^2$-BN sheets starts close to the central nitrogen vacancy and extend to the edge of the monolayer. The boundaries have a strongly disturbed symmetry of covalent bonds between the atoms [34] and can be categorized as extended defects originating from a nitrogen vacancy.

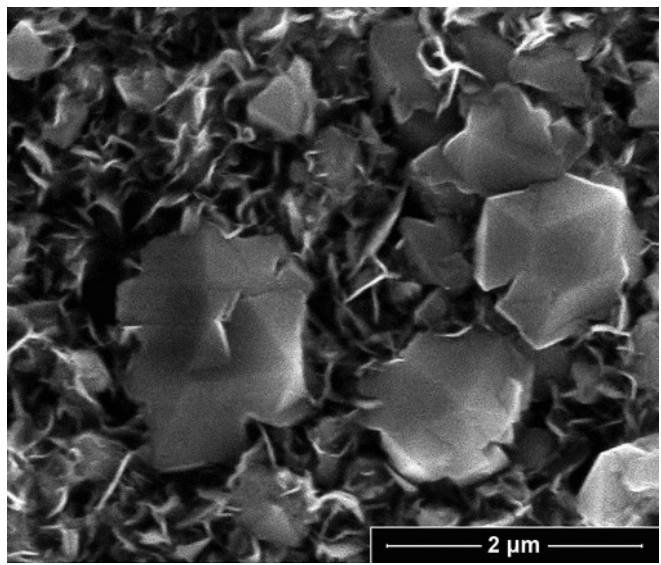

**Figure 9.** SEM image of a BN layer grown under HAHP conditions at T = 1050°C. The average thickness was estimated from reflectometry to 800nm.

Assuming the possibility of spatial deformation, by nitrogen vacancies, of free-standing $sp^2$-BN sheets, the density of such defects can be estimated to be larger than $10^8$ $cm^{-2}$ in the layer shown on Fig.9. Consequently, in the case of the material grown under LALP conditions, where the nitrogen vacancies density should be much higher, the quantity of folding points should be higher too. This conjecture can alternatively be taken to explain the dense morphology of the LALP material. The small grain sizes could be the result of a multiple folding of the hBN flakes.

The role of the TEB flow for the BN synthesis was precisely investigated for both growth conditions, HAHP and LALP. The growth rate was measured as a function of TEB flow with growth temperature fixed to T=1150°C. The results, shown on Fig.10, were obtained using multi-stage processes, with initially high TEB flow, consecutively decreased at subsequent stages. This ensured the fast creation of polycrystalline layers and later growth on composite surfaces similar to these shown in Fig.7 and Fig.8.

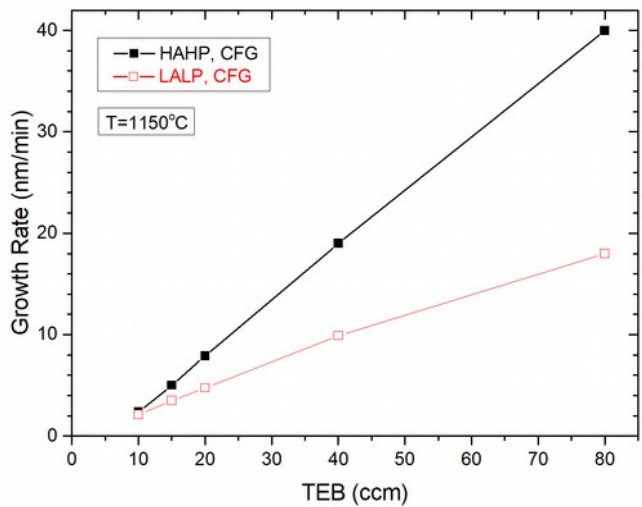

**Figure 10.** Growth rate dependence on TEB flow for two regimes of extreme conditions. Lines are guides to the eye only.

Both lines (Fig.10) are almost straight, with a small negative deviation for the lower TEB flows, which can be the result of TEB (i.e. boron ad-atoms) desorption from the layer surface. Ignoring this small effect and assuming a linear relation between TEB flow and BN growth rate, we can conclude that there is no influence of the TEB amount on the chemical synthesis efficiency, understood as the growth rate to TEB flow ratio. Exactly the same character of the dependency for HAHP and LALP conditions shows, that TEB plays always a passive role in the BN synthesis.

### 3.3 CFG processes limited by chemisorption (2D growth)

For processes initiated on sapphire substrates with relatively low TEB flow a self-limiting growth is observed. For example, the HAHP growth with TEB = 20 ccm, at T = 1300°C, did not yield observable results for reflectometry, for more than 60 minutes. According to the relationship illustrated in Fig.10, a growth rate of about 10 nm/min could be expected even at a lower temperature of T=1150°C for growth on already existing polycrystalline BN layers. After this long delay (> 60 min), growth became faster, reaching ~3 nm/min yielding a 3D porous layer. Ex-situ investigations of samples grown for 60 minutes reveal however an existence of 2D layers with thicknesses of about 5 nm. A SEM image of such a layer is presented on Fig.11.

Four types of objects are visible in the picture. Large, dark, round areas can be ascribed to local exfoliation of the layer. A mesh of dark lines originated from wrinkles resulting from the difference between the coefficients of thermal expansion of the layer and the substrate [22]. The black spots are hexagonal pits in the layer, observed also by AFM. The last visible type of objects are white and sharp-shaped, tens of nanometre long, free-standing BN flakes, similar to that dominating the layer shown in Fig.8. These stochastically nucleated flakes were created during the growth and their quantity increased with process time.

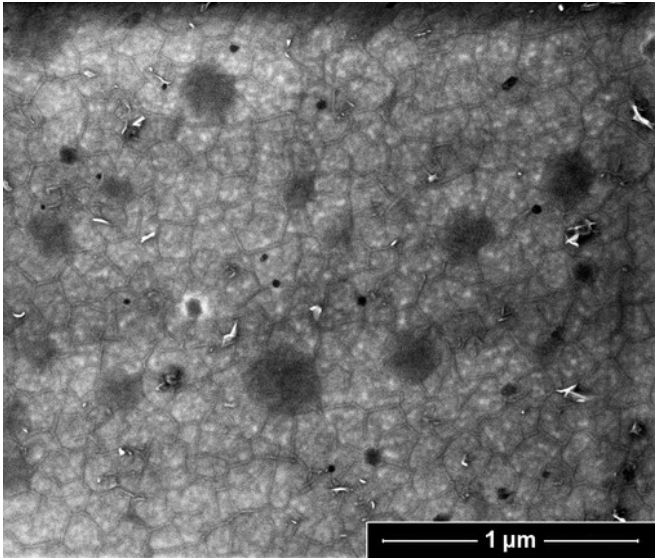

**Figure 11.** SEM image of a 5 nm thick BN layer (sample C) grown under HAHP conditions at T = 1300ºC with TEB = 20 ccm.

The flat morphology of the layer indicates an ordered stacking of sp²-BN sheets. Additionally it shows, that nitrogen vacancies, certainly created during the growth, cannot spatially deform the growing monolayer, if it is lying on the substrate. Van der Waals interactions between the sheets and the substrate prevent such deformation.

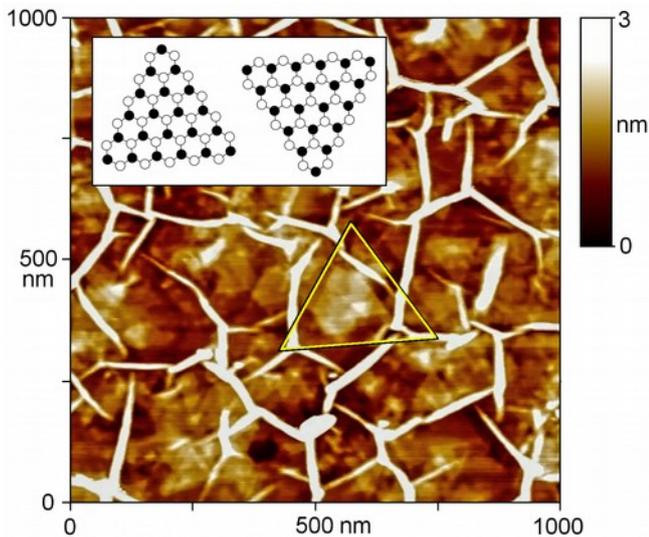

**Figure 12.** AFM image of a 5 nm thick BN layer grown under HAHP conditions at T = 1300ºC with TEB = 10 ccm. Triangular sp²-BN islands are visible between the ~3 nm high wrinkles. The inset shows the suggested surface reconstruction order of two subsequent mono-layers.

XRD and Raman spectroscopy confirm a highly ordered growth of the layer (see Fig.14 and Fig.15). However the most interesting information, about the growth character, can be found using atomic force microscopy (AFM). The image of the surface shown in Fig.12 is dominated by the mesh of wrinkles with heights of about 2-3 nanometers. Between the folds, however, steps with well-defined straight edges and heights of about 0.3 nm, can be observed. These steps, which can be up to 200 nm long, sometimes cross the folds and are arranged along only three directions and reveal triangularly shaped plateaues of single sp²-BN sheets. The ordering of the orientation of the triangles must be forced by the symmetry of the sapphire substrate, confirming true heteroepitaxial growth. Two directions of the orientations of triangles (see inset in Fig.12) reflect the sequence of consequent sp²-BN sheets at hBN crystal.

Similar triangles of monolayer BN islands were observed for different growth techniques and conditions. The shape of sp²-BN islands depends on the thermodynamic conditions, especially on the availability of nitrogen and boron (III/V ratio). [35,36]. The straight shapes and orientation of the atomic step lines in Fig.12 can be explained by the assumption that the edges of sp²-BN sheets are terminated by only one type of atoms – according to the sketch shown in the inset of Fig.12. Under HAHP condition, the edges have to be nitrogen-terminated.

The straight character of the atomic steps implies a high anisotropy of growth, with low efficiency in directions perpendicular to the steps and high efficiency in parallel directions. After establishing the triangular shape, the only dissimilar edge sites, on the whole sp²-BN sheet are the apexes of the triangle and possible crystallographic extended defects disturbing atomic reconstruction of the edges. These dissimilar sites should be responsible for further growth, in analogy to some extended defects in classical 3D crystal growth.

The described mechanism explains the observed fast suppression of 2D growth. Initially, the layer has to nucleate on the substrate as a pattern of separated, nanometric seeds. Each seed can grow relatively fast, by lateral expansion, until coalescence with the neighbours. After formation of a continuous, flat surface, the quantity and accessibility of the sp²-BN edges and apexes decreases radically and chemisorption practically stops. A simple prolongation of the process can result in nucleation of disordered seeds and the transition to 3D growth. The rate of such a transition strongly depends on the TEB saturation.

The above argumentation supplements and partially replaces the theory about a fast diminishing, catalytic influence of the substrate on the growth. Our results indicate that the growth suppression rate is correlated to the flattering of the layer surface. Consequently, self-termination of the growth can appear not only for monolayer films, but even much later, for layers with thicknesses of several nanometers. A catalytic influence of the substrate, as for example commonly observed on metals, should be negligible already for the deposition of the second monolayer. The described edge-termination mechanism is not dependent on deposition techniques and can act in wide range of conditions, until the amount of nitrogen ad-atoms is large enough to stabilise the sp²-BN edges. In order to resume and sustain 2D growth, changes of the boron chemisorption probability are required, which is indirectly a function of ammonia partial pressure.

### 3.4 FME growth

The FME method imposes an inherent limitation on the maximal possible growth rate. Given the fact that sp²-BN sheets are made of both, boron and nitrogen atoms, the amount of TEB, injected in a single pulse, should deliver material for not more than a fraction of a monolayer. Excess of boron quickly leads to problems with stoichiometry and disordered nucleation. On the other hand, the inertia of the control of the gas flows in an MOVPE equipment limits the shortest time of the single pulse to about 1 second. As result, a single atomic sheet can be grown within about 20 seconds, which corresponds to 1 nm/min only.

The efficiency of FME growth, as a function of the growth temperature, was investigated under HAHP conditions. The results shown in Fig.6 confirm, that the relationship character is exactly the same as for CFG at HAHP conditions. The result indicates, that in FME mode, the synthesis takes place mostly during ammonia-rich stages.

The experiments were performed with TEB = 15 ccm and a switching sequence (1.5; 0.0; 3.0; 0.5;), where the numbers indicate the time period (in seconds) of the TEB pulse, first interrupt, $NH_3$ pulse and second interrupt, respectively. An important observation is, that the length of the first interrupt, between TEB and $NH_3$ pulses, has no significant influence on the growth efficiency. The second interrupt however, responsible for the removal of ammonia from the reactor, plays a much more important role. The liquidation of this half-second long interrupt results in a threefold reduction of the growth efficiency. This finding confirms, that even residual amounts of ammonia uphold the passivation of the surface and block the subsequent chemisorption of boron ad-atoms.

According to our line of arguments, the positive results of FME in BN growth can be explained as activation of the boron chemisorption on $sp^2$-BN sheet edges caused by a periodic removal of ammonia (nitrogen) from the layer surface. This lack of nitrogen can facilitate reconstruction of atomic bonds under boron-rich conditions, but more probably enables corrosion of the edges of the $sp^2$-BN sheets, similar to what has been observed after annealing BN films in hydrogen [37]. Such an aggressive mechanism could be described as "one step back to take two steps forward" - after each corrosion (roughening) stage, the edges can be reconstructed a bit further during the ammonia-rich stage, before the next cycle starts. Periodic repetition of the switching sequence leads to a slow, but persistent growth.

The FME method stabilises and sustains 2D growth, nominally enabling production of layers with a thickness of hundreds of nanometers. In practice, the thickness of layers grown is such a way is still limited by a stochastic creation of 3D grains. The quantity of 3D grains increases with time of growth and finally dominates the surface. This process was the reason for the slight decrease of reflectivity signal with time shown in Fig.3. The probability of 3D nucleation strongly depends on the growth parameters and is subject to further optimisation. Fig.13 shows a SEM image of the layer grown using the FME mode under conditions similar to the CFG method for the sample shown on Fig.11. The sample grown by the FME method, for the same amount of time, reached seven times larger thickness, about 35 nm. The quantity and size of 3D grains on the surface remained similar, but the shape is rather round and grainy. The difference in shape can be ascribed to the periodic nitrogen-lack and non-stoichiometric composition. The overall cleaner appearance of the FME layer results mostly from the larger thickness.

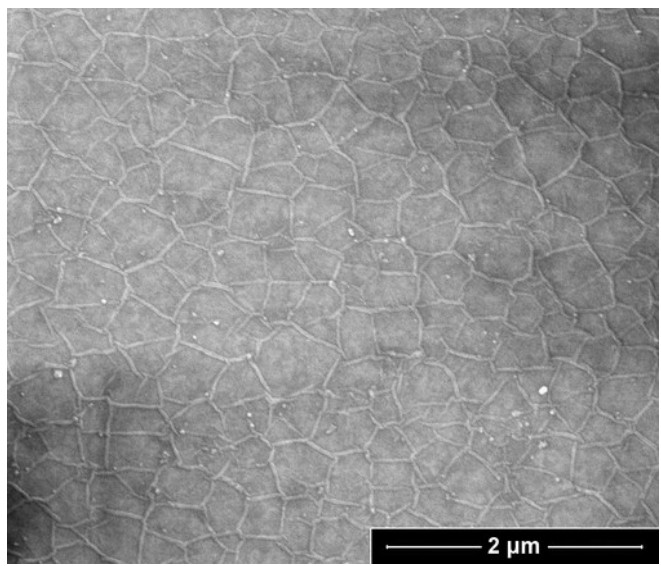

**Figure 13.** SEM image of a BN layer (sample D) grown using the FME mode at T = 1265°C. The layer thickness was 35nm.

### 3.5 Properties of the layers

As shown, the methods (CFG or FME) and conditions determine the efficiency of growth and morphology of BN layers. Other physical properties are presented in the current sub-section for the same samples presented in Fig.7,8,11,13. The basic information about the samples is summarised in Table 1.

| Sample | Growth method | Growth T (°C) | TEB flow (ccm) | Layer structure | Thick-ness (nm) |
|--------|--------------|---------------|----------------|-----------------|-----------------|
| A | CFG-LALP | 1050 | 40 | 3D | 380 |
| B | CFG-HAHP | 1050 | 40 | 3D | 380 |
| C | CFG-HAHP | 1300 | 20 | 2D | 5 |
| D | FME-HAHP | 1265 | 15 | 2D | 35 |

**Table 1.** List of the samples used in the text to illustrate the relationship between growth conditions and the properties of the layers.

The structural properties visible on the SEM and AFM images are confirmed by X-ray measurements (Fig.14). 2theta/omega scans yield a very weak diffraction signal from polycrystalline layers and clear signal from continuous 2D type layers. The thickness estimated from the FWHM (Full Width at Half Maximum) of the XRD peak, are typically lower than the value calculated from optical reflectometry. The deviation of the XRD results is a simple consequence of the layer deformation appearing after the growth – the mesh of wrinkles and partial exfoliation from the substrate. The layer thickness can be better estimated by the XRR technique, which does not take into account any specific crystallographic structure. XRR thickness estimations are in much better agreement with optical reflectometry.

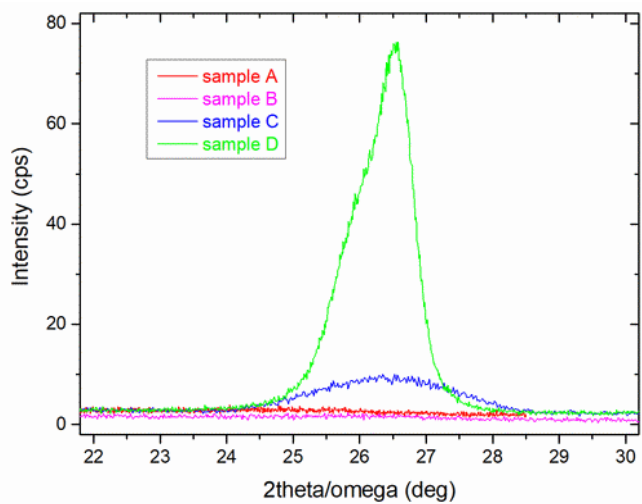

**Figure 14.** XRD 2theta/omega scans of BN layers grown under different conditions (see Tab.1). The scan for sample B is slightly shifted vertically for better visibility. Sample A and B show a barely visible diffraction signal (not distinguishable on this scale). The signal of sample C is broadened due to the low thickness of the layer.

Micro-Raman spectroscopy was performed as a part of the routine characterisation for every grown layer. The samples were excited using a laser with a wavelength of 532 nm. A clear Raman signal, at about 1368 $cm^{-1}$, which can be ascribed to the $sp^2$-BN structure [38], was observed for all samples (3D and 2D) grown under HAHP conditions (Fig.15). However, Raman spectroscopy was often obstructed by a background PL signal. The intensity of the background signal varied significantly between different samples grown with different parameters. In the case of samples grown under LALP conditions, the background intensity was too high to separate any Raman lines from the PL background.

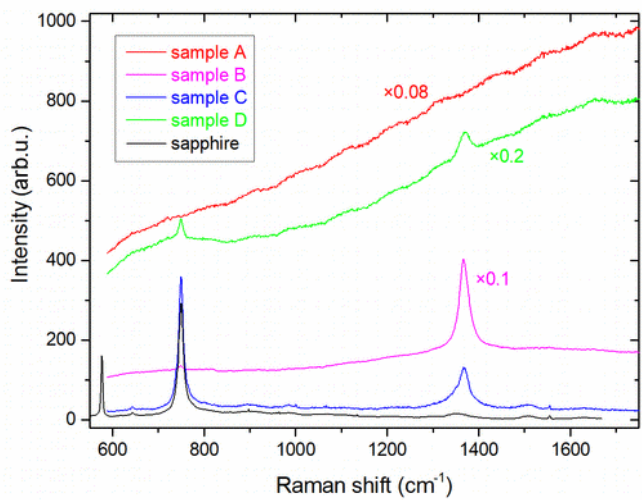

**Figure 15.** Raman spectroscopy spectra of BN layers grown under different conditions (see Tab.1). Sample A was excited with laser power reduced to 5%.

The presence a $sp^2$-BN structure in the layers grown under LALP conditions was, however, directly confirmed by FT-IR reflectometry measurements. Fig.16 presents reflection spectra of an area of 50 μm x 50 μm. A peak at around 1368 $cm^{-1}$, observed for all samples, corresponds to the TO branch of the in-plane $E_{1u}$ hBN phonon mode. Please note that for the system of a thin film of hBN on sapphire the reflectivity rather resembles the shape of absorption spectra and a clear Reststrahlen band, as in the case of bulk samples, is not expected. The additional feature visible at around 800 $cm^{-1}$ can be ascribed to the out-of-plane $A_{2u}$ modes. These modes are only visible for samples showing 3D morphology, enabling the observation of such phonon modes.

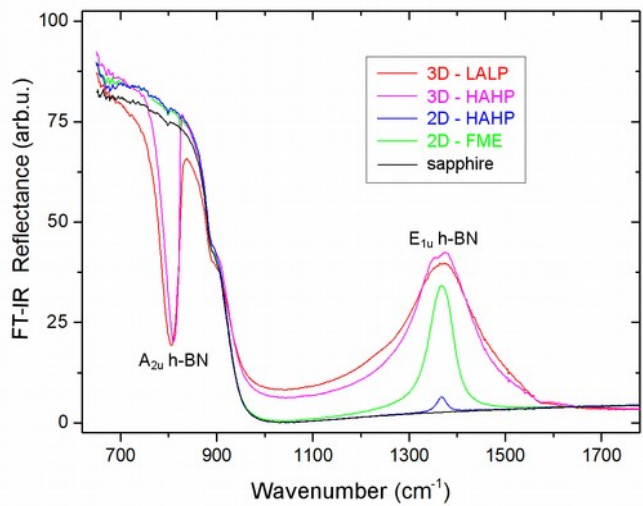

**Figure 16.** FT-IR reflectometry spectra of BN layers grown under different conditions (see Tab.1).

Investigation of the PL itself was performed by using another measurement system, but the same wavelength for excitation as for Raman spectroscopy. The results are presented in Fig.17. The spectra were not normalized with respect to the thickness of the layers. Because of the low absorption of visible light, it can be assumed, that PL is excited in the whole thickness of the layers and that the PL intensity can be dependent on the layer thickness.

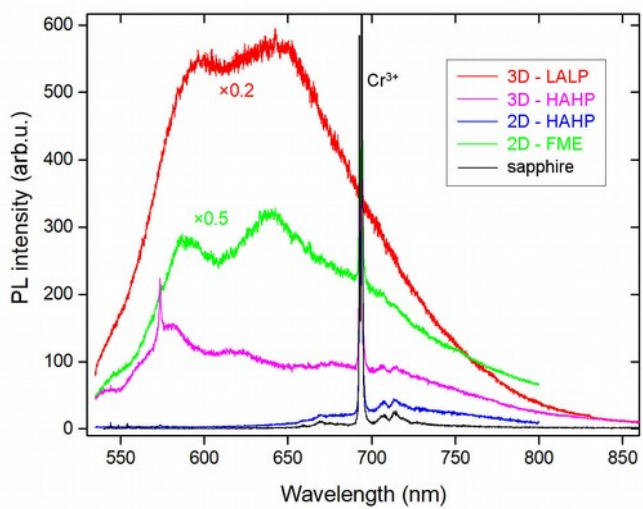

**Figure 17.** Room temperature photoluminescence spectra of the BN layers grown under different conditions (see Tab.1). The narrow line at 694 nm comes from the sapphire substrate (Cr3+). The line at about 575 nm is due to Raman scattering on $sp^2$-BN.

Taking into account the method and the growth conditions, we can directly correlate the integrated intensity of the PL with ammonia accessibility during the growth. The weakest PL was observed for the thinnest layer (sample C), grown with large ammonia amount and at the highest temperature. Even normalisation in terms of thickness does not change this result. The strongest PL signal was observed on sample A, grown with low ammonia amount and at relatively low temperature, which additionally lowers the atomic nitrogen accessibility. However by taking into account the lower thickness of sample D, it has to be agreed, that the most efficient luminescence comes from the layer grown at HAHP conditions, at high temperature but in FME mode.

The exact nature of the defects, responsible for the mid-gap luminescence, remains unrecognised [39]. The ammonia insufficiency can result intercalation of carbon and hydrocarbon radicals between sp²-BN sheets, carbon substitution at nitrogen sites or even the creation of nitrogen vacancies. Regardless of the nature of the defects giving rise to mid-gap PL, it is a very important observation that the FME method, resulting in 2D growth, most probably generates the highest quantity of such defects. This observation is in very good agreement with our picture of an aggressive corrosion of the atomic steps on hBN surface during ammonia-lack stages.

## 4. Conclusions

The interpretation of the presented experimental results obtained in a wide parameter range allows us to propose a consistent description of boron nitride MOVPE growth with ammonia and triethylboron precursors. Several specific phenomena were observed enabling a better understanding of the mechanisms governing sp²-BN layer creation.

The dependency of synthesis efficiency on ammonia flow, reactor pressure and process temperature contradict the widespread opinion, that parasitic gas-phase reactions strongly influence the BN MOVPE synthesis. Experiments performed in a wide parameter range reveal, that depending on ammonia accessibility, two different ammonia decomposition mechanisms can be activated. Low ammonia flow at low reactor pressure enables the heterogeneous pyrolysis of ammonia, leading to the synthesis of boron nitride even at 550ºC. This finding is very important, since it enables BN growth on substrates not sustaining high temperatures, like GaN, which might open the pathway for complex III-nitride epitaxy including hBN. The synthesis efficiency under these conditions depends linearly on the temperature, which suggests that the growth is limited by physical processes, like diffusion and desorption. A high ammonia flow at high reactor pressure (HAHP) enables the synthesis only above 950ºC, when the pyrolysis of ammonia provides nitrogen atoms. The efficiency increases fast with temperature and saturates above 1200ºC, indicating a complete ammonia decomposition. Both types of conditions enable a fast growth of 3D, porous material, consisting of free-standing flakes. The flakes size and shapes are influenced by ammonia amount, which shows, that nitrogen vacancies can spatially deform free-standing sp²-BN sheets.

Ordered 2D growth can be obtained at HAHP conditions using low TEB flow. But, after the fast formation of a layer with a thickness of a few nanometers, the growth is suppressed. The cause of this obstruction is identified to be the nitrogen-rich reconstruction of the atomic steps, blocking chemisorption of the boron ad-atoms. A direct and observable consequence of such a reconstruction are triangle-shaped sp²-BN islands on the surface of the layer. This description is more universal, than the theory about a catalytic influence of the substrate, and explains the self-termination of thicker layers, at high temperatures. The most important advantage of our description is the possibility to explain very effective 3D growth and the mechanism behind the successful 2D growth with the FME method.

Flow modulation epitaxy sustains 2D growth by a perturbation of the surface reconstruction. A periodic corrosion of the atomic steps, occurring during the nitrogen-poor stage, opens up the possibility of fast regrowth followed by a stabilisation during the subsequent, nitrogen-rich stage. Sequential roughening and reconstructing of the atomic steps enables the growth of thicker layers.

According to the above presented considerations, we find that the growth of hBN by MOVPE is ruled by very peculiar mechanisms, originating strictly from the sp² structure of hBN. In particular, the amount of ammonia influences diffusion and chemisorption of boron ad-atoms, changing synthesis efficiency and the layer morphology. Additionally it is responsible for the creation of optically active defects, observed in visible light spectrum. The consistent picture of hBN growth by MOVPE developed in this report lays the foundation for a further efficient, deterministic optimisation process which is indispensable to meet the requirement of prospective applications.